\documentstyle[epsfig,a4,12pt]{article}
\begin{document}
%
\newcommand{\x}{\cdot}
\newcommand{\ra}{\rightarrow}
\newcommand{\pom}{\mbox{${\rm \cal P}$omeron}}
\newcommand{\flux}{\mbox{$F_{{\cal P}/p}(t, \xi)$}}
\newcommand{\ap}{\mbox{$\bar{p}$}}
\newcommand{\pap}{\mbox{$p \bar{p}$}}
\newcommand{\SPS}{\mbox{S\pap S}}
\newcommand{\xp}{\mbox{$x_{p}$}}
\newcommand{\sumet}{\mbox{$\Sigma E_t$}}
\newcommand{\mpr}{\mbox{${m_p}$}}
\newcommand{\mpi}{\mbox{${m_\pi}$}}
\newcommand{\rs}{\mbox{$\sqrt{s}$}}
\newcommand{\rsp}{\mbox{$\sqrt{s'}$}}
\newcommand{\rsps}{\mbox{$\sqrt{s} = 630 $ GeV}}
\newcommand{\lum}{\mbox{$\int {\cal L} {dt}$}}
\newcommand{\T}{\mbox{$t$}}
\newcommand{\abt}{\mbox{${|t|}$}}
\newcommand{\di}{\mbox{d}}
\newcommand{\HS}{\mbox{$xG(x)=6x(1-x)^1$}}
\newcommand{\sigdifjets}{\mbox{$\sigma_{sd}^{jets}$}}
\newcommand{\sigpomjets}{\mbox{$\sigma_{{\cal P}p}^{jets}$}}
\newcommand{\sigdiftot}{\mbox{$\sigma_{sd}^{total}$}}
\newcommand{\sigpomtot}{\mbox{$\sigma_{{\cal P}p}^{total}$}}
\newcommand{\sigpomzero}{\mbox{$\sigma_{{\cal P}p}^o$}}
\newcommand{\dsig}{\mbox{${d^2 \sigma }\over{d \xi dt}$}}
\newcommand{\alamb}{\mbox{$\overline{\Lambda^{\circ}}$}}
\newcommand{\lamb}{\mbox{$\Lambda^{\circ}$}} 
\newcommand{\peetee}{\mbox{${ p_t}$}}
\newcommand{\PRET}{\mbox{\Proton-\sumet}}
\begin{titlepage}
\vspace{3cm}
\begin{flushright}  
{17 March, 1998}
\end{flushright}

\vspace{4ex}
\begin{center}
{
\LARGE\bf
\rule{0mm}{7mm} Evidence for $\xi$-- and \T -dependent damping \\
\rule{0mm}{7mm} of the \pom\ flux in the proton \\ 
}
\vspace{11 ex}
Samim Erhan and Peter E. Schlein \\
\vspace{3.0mm}
University of California$^{*}$, Los Angeles, California 90095, USA. \\
\end{center}
\vspace{11 ex}
\begin{abstract}
We show that a triple-Regge parametrization of inclusive single diffraction
agrees with the data in the following two domains:
(a) $\xi > 0.03$ at all \T , (b) $|t| > 1$~GeV$^2$ at all $\xi$.
Since the triple-Regge parametrization fails when applied to the full 
$\xi$--\T\ range of the total single-diffractive cross section, 
we conclude that damping occurs only at low--$\xi$ and low--$|t|$.
We give a (``toy'') parametrization of the damping factor, D($\xi$),
valid at low-$|t|$, which describes the $d\sigdiftot /dt$ data at the ISR
and roughly accounts for the observed $s$--dependence of \sigdiftot\
up to Tevatron energies.
However, an effective damping factor calculated for the CDF fitted function 
for $d^2\sigdiftot /d \xi dt$
at $\sqrt{s} = 1800$~GeV and $|t| = 0.05$~GeV$^2$, 
suggests that, at fixed-$\xi$, damping increases as $s$ increases.

We conjecture that, in the regions where the triple-Regge formalism
describes the data and there is no evidence of damping, 
factorization is valid and the \pom -flux-factor
may be universal.
With the assumption that the observed damping is due to multi-\pom\ exchange,
our results imply that the recent UA8 demonstration that the
effective \pom\ trajectory flattens for $|t| > 1$~GeV$^2$ is
evidence for the onset of the perturbative 2-gluon pomeron.
Our damping results may also shed some light on the self-consistency of
recent measurements of hard-diffractive jet production cross sections
in the UA8, CDF and ZEUS experiments.

\end{abstract} 
\vspace{2 ex}
\begin{center}
submitted to Physics Letters B \\
\end{center}
\vspace{17 ex}
\rule[.5ex]{16cm}{.02cm}
$^{*}$\ Supported by U.S. National Science Foundation Grant PHY94--23142\\
\end{titlepage}
\setlength{\oddsidemargin}{0 cm}
\setlength{\evensidemargin}{0 cm}
\setlength{\topmargin}{0.5 cm}
\setlength{\textheight}{22 cm}
\setlength{\textwidth}{16 cm}
\setcounter{totalnumber}{20}
\clearpage\mbox{}\clearpage
\pagestyle{plain}
\setcounter{page}{1}

\section{Introduction}
\label{intro}
\indent

The inclusive (inelastic) production of beam--like particles, known as single
diffraction, as in:
\begin{equation}
\ap \, \, + \, \, p_i \, \,  \ra \, \, X \, \, 
+ \, \, p_f \, \, \, \, \, \, \, \, \, \, \, + \, c.c.
\label{eq:dif}
\end{equation}
and its analogous $pp$ and ep interactions, 
presents one of the most interesting phenomena in strong interaction physics. 
An observed ``rapidity gap" (absence of particles in a range of rapidity) 
between $X$ and $p_f$ in the final state
signifies that the entire (color singlet) residual momentum of the proton,
with beam momentum 
fraction, $\xi = 1 - \xp$, participates in the
interaction between it and the second beam particle. This effect is
described in terms of the exchange of the \pom\ Regge 
trajectory\cite{reviews},
which embodies the idea of ``factorization". The momentum transfer, \T , and 
the beam momentum fraction, \xp , of
the final state proton, $p_f$, ``tag'' the corresponding parameters of
the exchanged \pom .

Since $\xp \sim 1$ is observed
to be the most likely beam momentum fraction of the final--state
$p_f$, correspondingly the most likely value of the \pom 's
momentum fraction in the proton, 
$\xi$, is near zero. 
Nonetheless, at current collider energies
the squared--invariant--mass of the system $X$ in Eq.~\ref{eq:dif}, 
$\, \, s' = \xi s$ to good approximation, 
can be quite large. This fact led to
a proposal\cite{is} to study hard scattering in such interactions, as a means
of determining if the \pom\ possesses an observable partonic structure.
The observation of the predicted hard scattering\cite{bonino,brandt,jetsig} 
supported
the notion that the \pom\ behaves like a quasi--real object inside
the proton with an effective \pom\ flux factor. 
An open question is
to what extent such a flux factor is universal; for example, 
is it independent of beam particle or center-of-mass energy,
or are there regions of phase space where factorization breaks down,
due to interference with other more complex phenomena 
(e.g. multiple-\pom -exchange) ?

One of the long--standing theoretical problems in high
energy hadronic interactions has been the understanding of $s$--channel
unitarization in \pom --exchange (diffractive) interactions.
Empirically, one finds that the total diffractive cross section,
\sigdiftot ,
in Reaction~\ref{eq:dif} and in the corresponding $pp$ interaction,
initially rises from threshold and tends to level off or
``flatten" at high energy\cite{dino}, 
whereas the dominant triple--\pom \cite{mueller} 
description of 
these processes (see below) continues to rise and soon exceeds the total $p\ap$
cross section. There is no built--in mechanism in the pure triple--\pom\
process to account for the observed flattening of \sigdiftot , and hence
avoid the violation of unitarity.

Figure~1 displays the problem~[8--17]. 
The $s$-dependence of the total cross section for React.~\ref{eq:dif},
\sigdiftot\,
is shown\footnote{It is conventional to quote \sigdiftot\ for 
$\xi_{min} <\xi < 0.05$,
because the experimental acceptance usually depends weakly on \xp\ in this 
region and because the integrated background from non-\pom\ exchange and other 
sources in this region is small enough to be neglected.}
for Feynman--$\xp > 0.95$ (or $\xi < 0.05$) 
of the final state proton or antiproton
(the domain where \pom --exchange is dominant).
\sigdiftot\ rises sharply from its threshold
at 11.3 GeV beam momentum  and gently levels off to $\sim$9~mb 
at the highest Fermilab energy. 
The solid curve in Fig.~\ref{fig:sigtot} is the triple-Regge prediction
discussed below. At high energies, it is in complete disagreement with
the measured cross section.

In the continuing theoretical efforts to satisfy $s$--channel
unitarity [18--22],
the words, screening, shadowing, absorption and damping are all 
used\cite{kaidalov} to
describe effects due to multiple \pom\ exchange
(two-\pom -exchange is also an important
component in understanding $pp$ elastic scattering\cite{dl_elastic} 
at low-$|t|$). These calculations have had varying degrees of success.
Goulianos has taken a more pragmatic approach\cite{dino} 
to satisfying unitarity
and suggested that the integral
of the \pom\ flux factor in a proton should saturate at 
unity above  $\rs \sim 22$~GeV.

In the present Letter, we find that damping is confined to the low--$\xi$, 
low--$|t|$ region.
We continue the analysis of the UA8 
Collaboration\cite{ua8dif} and demonstrate that there are regions 
either at larger $\xi$ or at larger \T , where the available data are
well described by the triple-Regge formula and therefore require no damping.

It is thus
clear that the damping function depends on both $\xi$ and \T .
We attempt to determine the $\xi$--dependence at low--$|t|$ of an ``effective" 
multiplicative damping factor 
which could account for the discrepancies between data and solid curve in
Fig.~\ref{fig:sigtot}. 
However, we call it a ``toy damping factor" for several reasons.
First, there are large gaps in the available data in Fig.~\ref{fig:sigtot}
and some inconsistencies,
therefore making it impossible to find a unique function.
Secondly, the processes which give rise to the observed damping may 
imply a breakdown of factorization,
in which case a simple universal damping factor may not exist
at low--$\xi$ and low--$|t|$.
Finally, there is some evidence that the nature of the $\xi$--dependence
may itself depend on $s$ at our highest energies.

\section{Triple--Regge phenomenology}
\label{tripleR}
\indent

We briefly summarize the relevant formula.
The Mueller--Regge expansion\cite{mueller} for the differential cross section 
of 
React.~\ref{eq:dif} is:
\begin{equation}
{{d^2\sigma_{sd}}\over{d\xi dt}}\, \, = 
\, \, \sum_{ijk} \, G_{ijk}(t) \x \xi^{1-\alpha_i(t) - \alpha_j(t)} 
\x (s')^{\alpha_k(0) - 1}
\label{eq:tripleR}
\end{equation}
where $\alpha_i (t)$ is the Regge trajectory for Reggeon $i$.
The sum is taken over all possible exchanged Reggeons.
The $G_{ijk}(t)$ are products of the various Reggeon--proton and 
triple--Reggeon 
couplings and the signature factors. 

There are two dominant terms in Eq.~\ref{eq:tripleR} 
at small $\xi$, namely $ ijk = {\cal P}{\cal P}{\cal P}$
and ${\cal P}{\cal P}{\cal R}$,
where the first term corresponds to the triple--\pom\ process, 
and the second corresponds to other non--leading, C=+ trajectories 
(e.g., $f_2$) in the \pom --proton interaction,
The ${\cal P}{\cal P}{\cal P}$ term increases with increasing $s'$, 
whereas the ${\cal P}{\cal P}{\cal R}$ term decreases with increasing $s'$.

Because the \pom\ is the highest--lying trajectory, 
when $i=j=\pom $, $1 - 2 \alpha$ is negative and
the differential cross
section increases sharply as $\xi \ra 0$. This corresponds to the empirical
observation that the most likely momentum fraction of the \pom\ in the
proton, $\xi$, is near zero. 
Thus, the sharp rise in the triple--Regge prediction
of \sigdiftot\ in Fig.~\ref{fig:sigtot} is due to
the kinematic 
fact that the minimum value of $\xi$ decreases with increasing $s$
as $\xi_{min} = s_{min}' / s$

For fitting to data, Eq.~\ref{eq:tripleR} has been rewritten\cite{ua8dif} as:
\begin{equation}
\label{eq:tripleP2}
{{d^2 \sigma_{sd}}\over{d \xi dt}} 
\, = \, [K \, F_1(t)^2  \, e^{bt} \, \xi^{1-2\alpha (t)}] 
\, \x \, \sigma_0 [(s')^{\epsilon_1} \, + \, \rm R \, \it (s')^{\epsilon_2}],
\end{equation}
where, 
\begin{itemize}
\item the left--hand bracket is taken as the \pom\ flux factor, \flux ,
and the right--hand bracket (together with the constant, $\sigma_0$) is 
the \pom --proton total cross section, \sigpomtot .
\item The two terms in \sigpomtot\ correspond to the $(s')^{\alpha_k (0) -1}$
factor\footnote{At very large $s'$, rescattering effects may lead to a 
logarithmic 
dependence on $s'$ as well as to other complications\cite{kaidalov}.} 
in the ${\cal P}{\cal P}{\cal P}$ and ${\cal P}{\cal P}{\cal R}$
terms in Eq.~\ref{eq:tripleR}.
Thus, \sigpomtot\ has a form similar to that of
real particle cross sections\cite{dl_tot}.
\item The products 
$K \sigma_0$ and $K \sigma_0 \rm R$ are, respectively, the values
of $G_{{\cal P}{\cal P}{\cal P}}(t)$ and $G_{{\cal P}{\cal P}{\cal R}}(t)$
at \T\ = 0. 
\item The \pom\ 
trajectory, $\alpha (t)$, 
has been shown\cite{ua8dif} to become relatively flat for
$|t| > 1$~GeV$^2$ (see next section); therefore
a quadratic term is added to the standard linear 
trajectory\cite{dl_elastic}, $\alpha (t)$ = 1.10 + 0.25 t + $\alpha'' t^2$.
The non--zero value of $b$ in $e^{bt}$ compensates for the presence of the 
quadratic component in the \pom\ 
trajectory\footnote{If, as Donnachie and Landshoff\cite{dl_dif} have suggested,
\sigpomtot\ depends on momentum transfer, that dependence would also
be absorbed into the $e^{bt}$ factor.}.
\item $|F_1(t)|^2$ is the standard Donnachie--Landshoff\cite{dl_elastic} 
form--factor.\footnote{$F_1(t)={{4 m_p ^2 - 2.8t}
\over{4 m_p ^2 - t}}\, \x \, {1\over{(1-t/0.71)^2}}$} 
Since it has never been shown to describe React.~\ref{eq:dif} at large \T ,
the $e^{bt}$ factor also serves as a possible correction.
Thus, the product, $|F_1(t)|^2 e^{bt}$, 
carries the \T --dependence
of the $G_{ijk}$  in Eq.~\ref{eq:tripleR} and is assumed to be the same for both
$G_{{\cal P}{\cal P}{\cal P}}$ and $G_{{\cal P}{\cal P}{\cal R}}$.
Physically, this means that the \pom\ has the same flux factor in
the proton, independent of whether the \pom --proton interaction proceeds
via \pom --exchange or Reggeon--exchange.
\end{itemize}

\section{Where is triple--Regge applicable ?}
\label{sec:applicable}
\indent

We already know from the information in Fig.~\ref{fig:sigtot}
that the dominant contribution to the total cross section, 
namely the data with small--$\xi$ and small--$|t|$,
are not described by the triple--Regge formalism;
a damping of the \pom\ flux with increasing $s$ is certainly 
required in this region.
However, we see no reason to suppose that
the same damping must apply to
the entire $\xi$--\T\ domain\footnote{\pom --exchange dominates out to
$\xi \sim 0.05$ and contributes significantly to $\xi \sim 0.1.$},
as proposed by Goulianos\cite{dino}. 
However, this issue can be resolved by using available data 
to determine if there
are regions in the $\xi$--\T\ plane
where the formalism does apply; that is, where damping is not required.

The UA8 collaboration has recently reported\cite{ua8dif} a 
(successful) simultaneous fit
of Eq.~\ref{eq:tripleP2} to their data on 
React.~\ref{eq:dif} at $\sqrt{s} = 630$~GeV and the extensive data sample
of the CHLM collaboration at the CERN--ISR with $\sqrt{s}$ = 23.5 and 30.5~GeV.
They use the values, $\alpha_k (0) -1$ = 0.10 and -0.32, obtained
for the \pom\ and $f/A2$ trajectories, respectively, in fits to
real--particle total cross sections\cite{dl_tot,cudell,dino2}.  

The four free parameters, 
$K\sigma _0$,  $\alpha'' , \, b \, \, \rm and \, \, \it R$,
are determined by fitting Eq.~\ref{eq:tripleP2}, plus an empirical background
function of the form, $A e^{ct} \xi^1$, to the 
combined ISR--UA8 data set in the 
range\footnote{For $\xi > 0.03$, there are no concerns about experimental 
resolution causing ``spill--over'' from the large peak at $\xi \sim 0$.}, 
$\xi = 0.03$--0.10. The fitted parameters are:
\vspace{3.0mm}
\begin{tabbing}
\hspace{5cm}\=$K\sigma _0$ \hspace{3mm}\= =\hspace{4mm}\=$0.72 \pm 0.10$ 
\hspace{5mm}\=mb GeV$^{-2}$\\
           \>$\alpha ''$   \> =\>$0.079 \pm 0.012$ \>GeV$^{-4}$\\
           \>$b$	   \> =\>$1.08 \pm 0.20$   \>GeV$^{-2}$\\
           \>$R$           \> =\>$4.0 \pm 0.6$\\    
\end{tabbing}
This value of $\alpha ''$ was independently
confirmed\cite{ua8dif} in fits to the $\xi$--dependence in the peak
region with $\xi < 0.03$.

These triple--Regge results can be used to predict
the total cross section of React.~\ref{eq:dif}, \sigdiftot ,
be integrating Eq.~\ref{eq:tripleP2}
over the entire \T --range, as well as for $\xi_{min} < \xi < 0.05$.
This yields the solid curve in Fig.~\ref{fig:sigtot} and illustrates
the discrepancy\cite{dino} with the experimental \sigdiftot .

Fig.~\ref{fig:dsdt035} from Ref.~\cite{ua8dif}
shows the ISR\cite{albrow} and UA8 data\cite{ua8dif}
data in the restricted region, $0.03 < \xi < 0.04$, 
where the small ($\sim$15\%) non--\pom --exchange background can be ignored
(the background is smaller than the size of the dots in the figure
and about the same magnitude as the systematic uncertainty in absolute
cross sections). 
The similarity of \dsig\ at $\xi = 0.035$ at both ISR and \SPS\ energies 
reflects the fact that $\sigpomtot (s')$ has nearly the same value at
both $\sqrt{s'}$ = 5 and 118 GeV.\footnote{This arises because, 
at fixed $\xi$ and \T\ in Eq.~\ref{eq:tripleP2}, \dsig\ is proportional to 
$\sigpomtot (\xi s)$.} 
The term, $(s')^{-0.32}$, in Eq.~\ref{eq:tripleP2}
makes this possible.
The solid curves in Fig.~\ref{fig:dsdt035} are fits 
to these data {\it without a background term}
and yield values
of the 4 parameters which are in excellent 
agreement\footnote{The same four parameters from this fit are
$(0.67 \pm 0.08)$, $(0.078 \pm 0.013)$, $(0.88 \pm 0.19)$ 
and $(5.0 \pm 0.6)$, respectively.} 
with those given above from the $\xi = 0.03-0.10$ fit,
thus lending credence to the relibility and stability of the fits.

We have found a second region in the $\xi$--\T\ plane, at small--$\xi$ but 
large--$|t|$, where the 
triple--Regge formalism also describes the ISR and UA8 data --- with no 
additional free parameters.
Fig.~\ref{fig:sigtothit} shows the high momentum--transfer part of \sigdiftot\ 
for $\xi_{min} < \xi < 0.05$ and the limited $|t|$--range, 1.0--2.0 GeV$^2$, 
plotted vs. $s$ for the ISR and UA8 data. 
The solid curve is the prediction of Eq.~\ref{eq:tripleP2} using the above 
parameters. 
The dashed 
curve is obtained by decreasing the $b$--parameter by 1$\sigma$
from its central value. 
We see that, in contrast with the situation for \sigdiftot , 
the triple--Regge formula, Eq.~\ref{eq:tripleP2}, 
accounts for the 
observed $s$--dependence of the total single diffractive cross section 
in the high--$|t|$ range, 1.0--2.0~GeV$^2$.
The different shapes of the curves in Figs.~\ref{fig:sigtot} 
and the solid curve in Fig.~\ref{fig:sigtothit} 
are due to the \T --dependence of the \pom\ trajectory.

We have thus demonstrated that 
damping depends on both $\xi$ and \T ; 
it only exists in the small--$\xi$, small--$|t|$ region, and is
not required by data away from that region --- either at larger
$\xi$ or at larger $|t|$. 
This could explain why CDF\cite{cdf} reports abnormally
large backgrounds in triple--\pom\ fits
to their (low--$\xi$, low--$|t|$)
data at \rs\ = 1800 GeV.  For example, at $\xi = 0.035$ 
(and $|t|$ = 0.05 GeV$^2$) where normally
15--20\% background is found, their fitted formula
corresponds to non--\pom --exchange backgrounds of 51\%.
Such a result can be expected if a 
$\xi$--dependent Damping factor is required, but is left out of the fit;
the fitted (large) 
background term compensates for the 
wrong $\xi$--dependence in the triple-Regge equation without damping.
Since CDF only reports the fitted functions, we compare our prediction
of \dsig\ with the sum of their ``signal" and ``background" terms. 
This sum corresponds to the solid bands in Fig.~\ref{fig:cdft};
the curves are the \dsig\ vs. \T\ predictions of 
Eq.~\ref{eq:tripleP2} at $\xi = 0.035$. We see that, at 1800 GeV, the
prediction agrees to within $1\sigma$ of the CDF result; at 546 GeV, there
is also good agreement at the lowest $|t|$--value, although their
fitted $t$-dependences at the two energies are not self-consistent. 
 
\section{Empirical determination of damping at small ($\xi$,\T) }
\label{determination}
\indent

The \T --dependence of the disagreement between triple--Regge and
the measured cross section is best seen by comparing the predictions with
the experimental values of $d \sigdiftot /dt$ plotted vs. \T . 
This is done in  
Figs.~\ref{fig:dsdtall} for eight ISR energies\footnote{Some of these data 
were obtained with unequal energies for the two beams} 
and in Fig.~\ref{fig:dsdtua8} at the \SPS --Collider\footnote{The UA4 points
come from two independent runs, one at high--$\beta$ and one at low--$\beta$
which allowed them to span most of the available \T --range.}. 

The dashed and dotted curves in Figs.~\ref{fig:dsdtall} and \ref{fig:dsdtua8}
are the (undamped) 
triple--Regge predictions for $d\sigma / dt$, calculated by integrating
Eq.~\ref{eq:tripleP2} over the range, $\xi_{min} < \xi < 0.05$ 
(for the dotted curve, $b$ is decreased by 1$\sigma$ from its central value). 

\begin{figure}
\begin{center}
\mbox{\epsfig{file=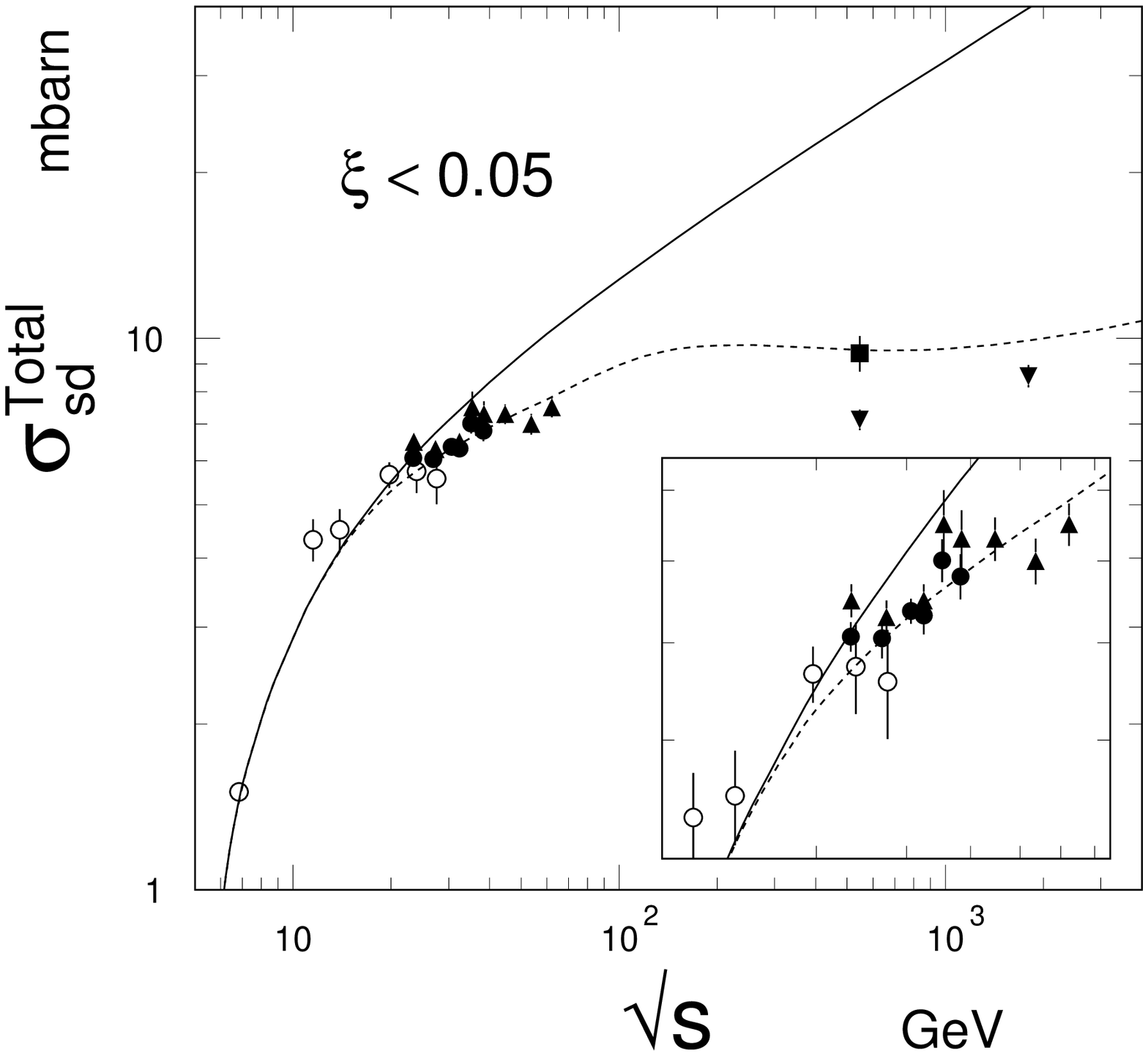,width=13cm}}
\end{center}
\caption[]{
\sigdiftot\ of $pp$ or $p\ap$ interactions (with $\xi < 0.05$) 
vs. $\sqrt{s}$ demonstrating the flattening of the
cross section with energy (a factor of two is included to account for
both hemispheres).
The insert is a blow--up of the ISR energy range.
The upper curve is the Triple--Regge prediction described in the text;
the dashed curve shows the consequence of 
multiplying it by the ``toy" damping factor discussed in the text.
The lowest energy points (open circles) are from bubble
chamber experiments [8--12]; followed by those from the ISR  
(solid circles\cite{albrowtot} and triangles\cite{armitagetot}),
the \SPS --Collider (solid square \cite{ua4dif1,ua4dif2}) 
and the Tevatron (inverted triangles \cite{cdf}).
}
\label{fig:sigtot}
\end{figure}

\begin{figure}
\begin{center}
\mbox{\epsfig{file=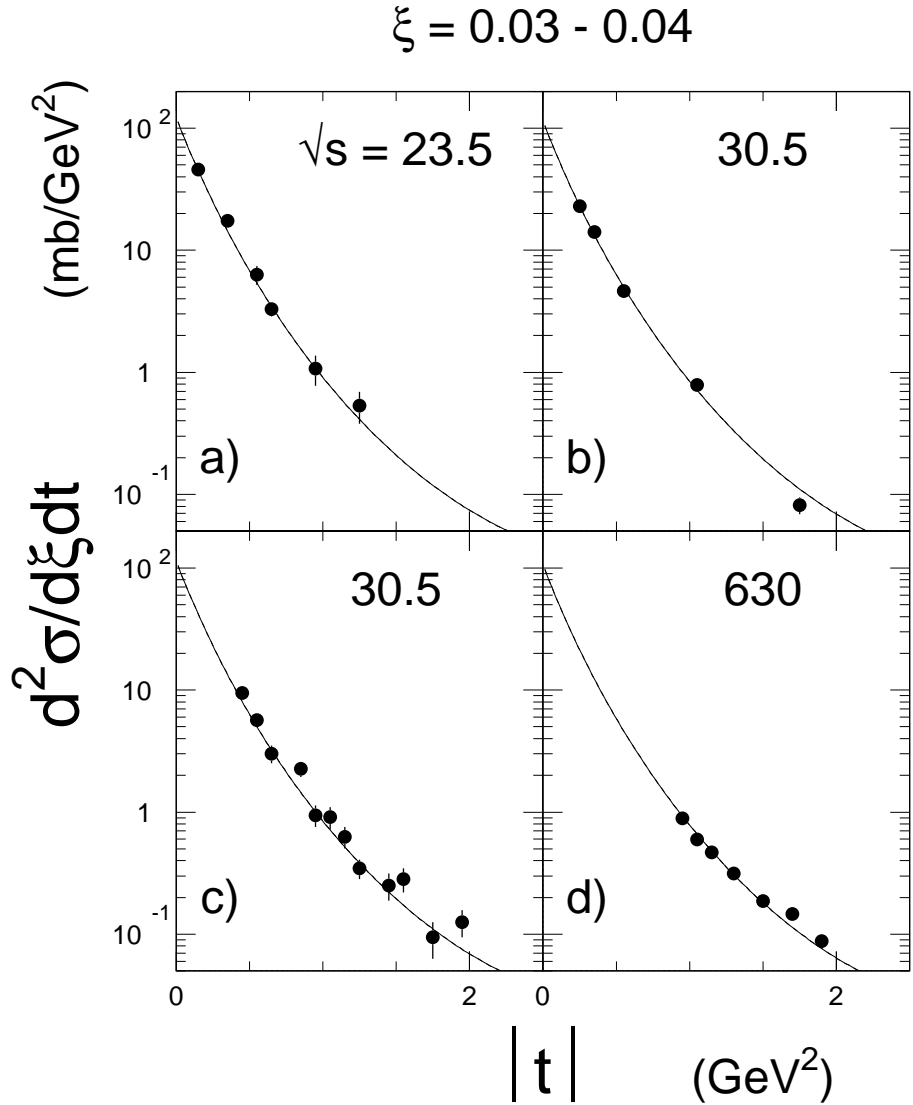,width=13cm}}
\end{center}
\caption[]{
Differential cross section, \dsig , vs \T , for 3 ISR 
measurements\protect\cite{albrow} and UA8\protect\cite{ua8dif}
(single--arm cross sections). 
The curves correspond to the fit described in the text~\protect\cite{ua8dif}.
The points are averages of data in the $\xi$--range 0.03--0.04
(the non--\pom --exchange background in the data points is about the same 
magnitude as the diameter of the dots).
}
\label{fig:dsdt035}
\end{figure}

\begin{figure}
\begin{center}
\mbox{\epsfig{file=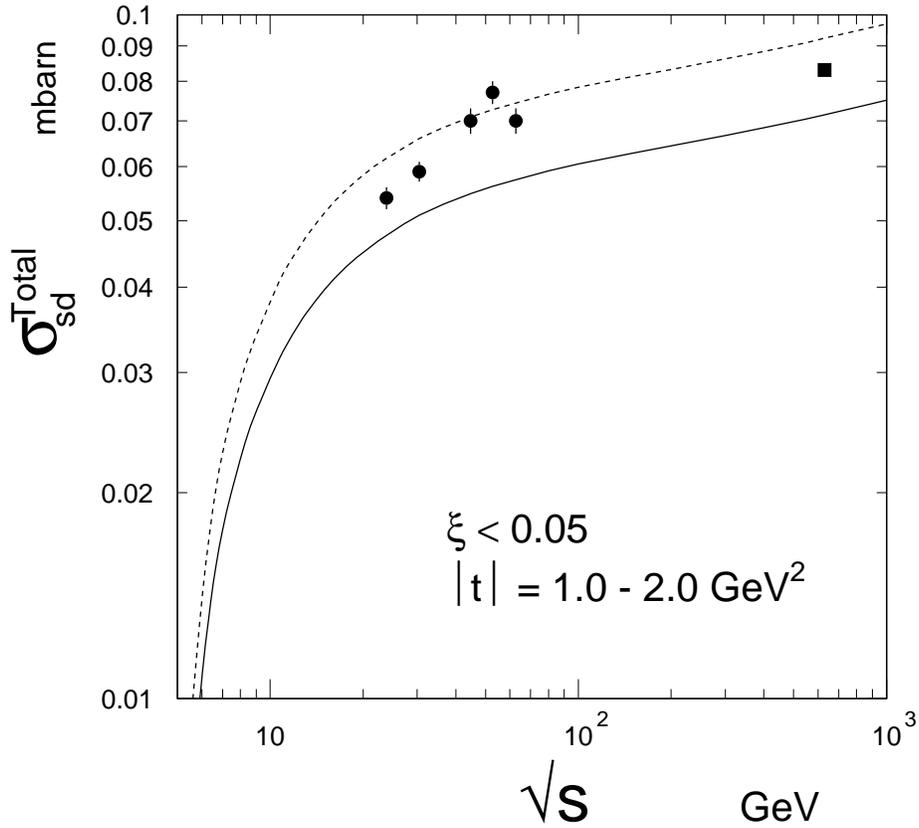,width=13cm}}
\end{center}
\caption[]{
\sigdiftot\ for $\xi < 0.05$ and $|t|$ = 1.0--2.0~GeV$^2$ vs. $\sqrt{s}$
(a factor of two is included to account for both hemispheres). 
The solid curve is the same Triple--Regge prediction used
in Fig.~\protect\ref{fig:sigtot} (where it is integrated over all \T );
the dashed curve is the same, but with
the ``$b$"--parameter decreased by 1$\sigma$ from its central value.
The lowest energy points (closed circles) are from the
ISR~\protect\cite{albrowtot}; 
the highest energy point (solid square) is from
the \SPS --Collider~\protect\cite{ua8dif}
}
\label{fig:sigtothit}
\end{figure}

\begin{figure}
\begin{center}
\mbox{\epsfig{file=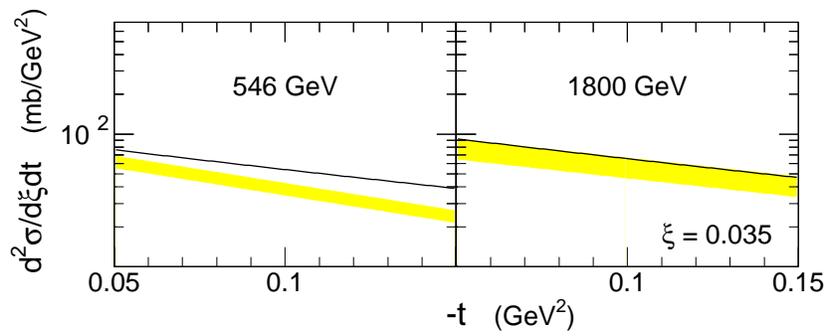,width=13cm}}
\end{center}
\caption[]{
Bands are the CDF differential cross sections at $\xi = 0.035$, calculated
from their fitted functions\cite{cdf} (single--arm cross sections); 
the band widths are $\pm$1$\sigma$ errors on their amplitudes 
(as explained in the text, their ``signal" and ``background"
are added together). 
The curves are from the same calculations used for the
curves in Fig.~\ref{fig:dsdt035}.
}
\label{fig:cdft}
\end{figure}

\begin{figure}
\begin{center}
\mbox{\epsfig{file=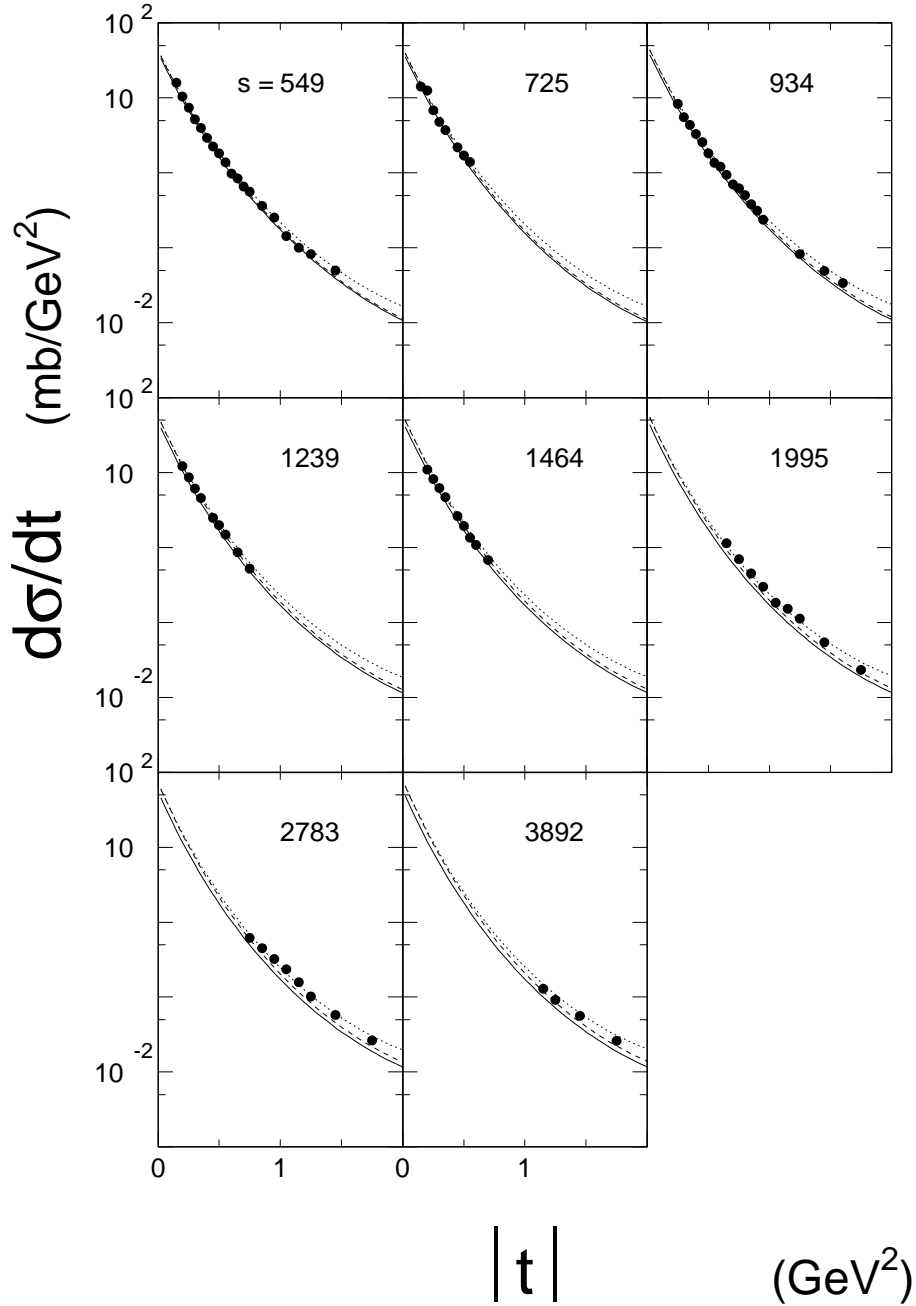,width=13cm}}
\end{center}
\caption[]{
$d\sigma / dt$ vs. $|t|$ (single--arm 
cross sections) with $\xi < 0.05$ at eight 
ISR\protect\cite{albrowtot,armitagetot} energies. The numbers shown in each
plot are their $s$ values (GeV$^2$).
The solid curves are the integrals of Eq.~\protect\ref{eq:tripleP2}
with damping included, using the parameters given in the text.
The dashed and dotted curves 
are calculated without damping; the dashed curve uses the central value
of $b$, while for the dotted curve, $b$ is  decreased
by 1$\sigma$ from its central value.
}
\label{fig:dsdtall}
\end{figure}

\begin{figure}
\begin{center}
\mbox{\epsfig{file=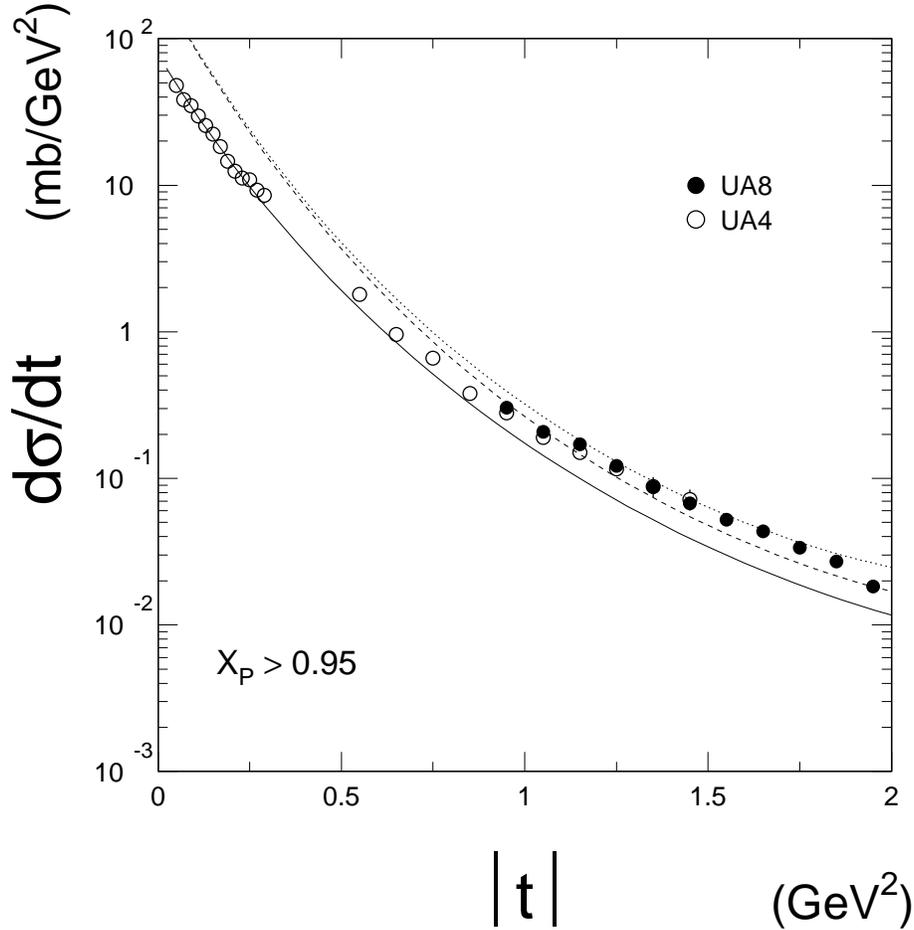,width=13cm}}
\end{center}
\caption[]{
Inclusive differential cross section for protons in React.~\protect\ref{eq:dif}
for $\xp > 0.95$, measured in experiment UA8 with \rs\ = 630 GeV.
and in experiment UA4\protect\cite{ua4dif1,ua4dif2} with \rs\ = 546 GeV
(single--arm cross sections; 
the integral is $4.7 \pm 0.35$ mb, or $9.4 \pm 0.7$ mb for \sigdiftot ).
The solid curve is the integral of Eq.~\protect\ref{eq:tripleP2}
with damping included, as explained in the text. 
The dashed and dotted curves 
are calculated without damping; the dashed curve uses the central value
of $b$, while for the dotted curve, $b$ is  decreased
by 1$\sigma$ from its central value.
}
\label{fig:dsdtua8}
\end{figure}

At the ISR energies, where the triple--Regge prediction 
only exceeds the data by about 10--15\% (see Fig.~\ref{fig:sigtot}), 
the differences
between dashed curves and data in Fig.~\ref{fig:dsdtall} are hardly noticable, 
because the dot sizes are roughly similar to the discrepancies. 
At $\rs = 630$~GeV,  however, the same effect is larger and highly visible.
At that energy, we see that there is a gradual transition from the low--$|t|$ 
region which dominates \sigdiftot , and where the experimental \sigdiftot\ is 
smaller than the (undamped) triple--Regge prediction, to the higher--$|t|$ 
region where the predictions agree with the data. 
This seems to be a smooth transition over the $|t|$--range, 0.5--1.0~GeV$^2$. 
The situation at the lower ISR energies in Figs.~\ref{fig:dsdtall} 
is similar but less pronounced.
We conclude that the discrepancies between predictions and data are confined to 
the low--$|t|$ region.

Since, as noted above, the calculated rise in \sigdiftot\
with increasing $s$ is due to the fact that, kinematically, 
$\xi_{min}$ decreases with increasing
$s$, it seems natural to introduce an empirical damping factor, $D(\xi)$, in
Eq.~\ref{eq:tripleP2} which suppresses small $\xi$--values;
{\it i.e., we strike at the ``heart'' of the problem}.
D($\xi$) will be unity everywhere except at small--$\xi$.

\begin{figure}
\begin{center}
\mbox{\epsfig{file=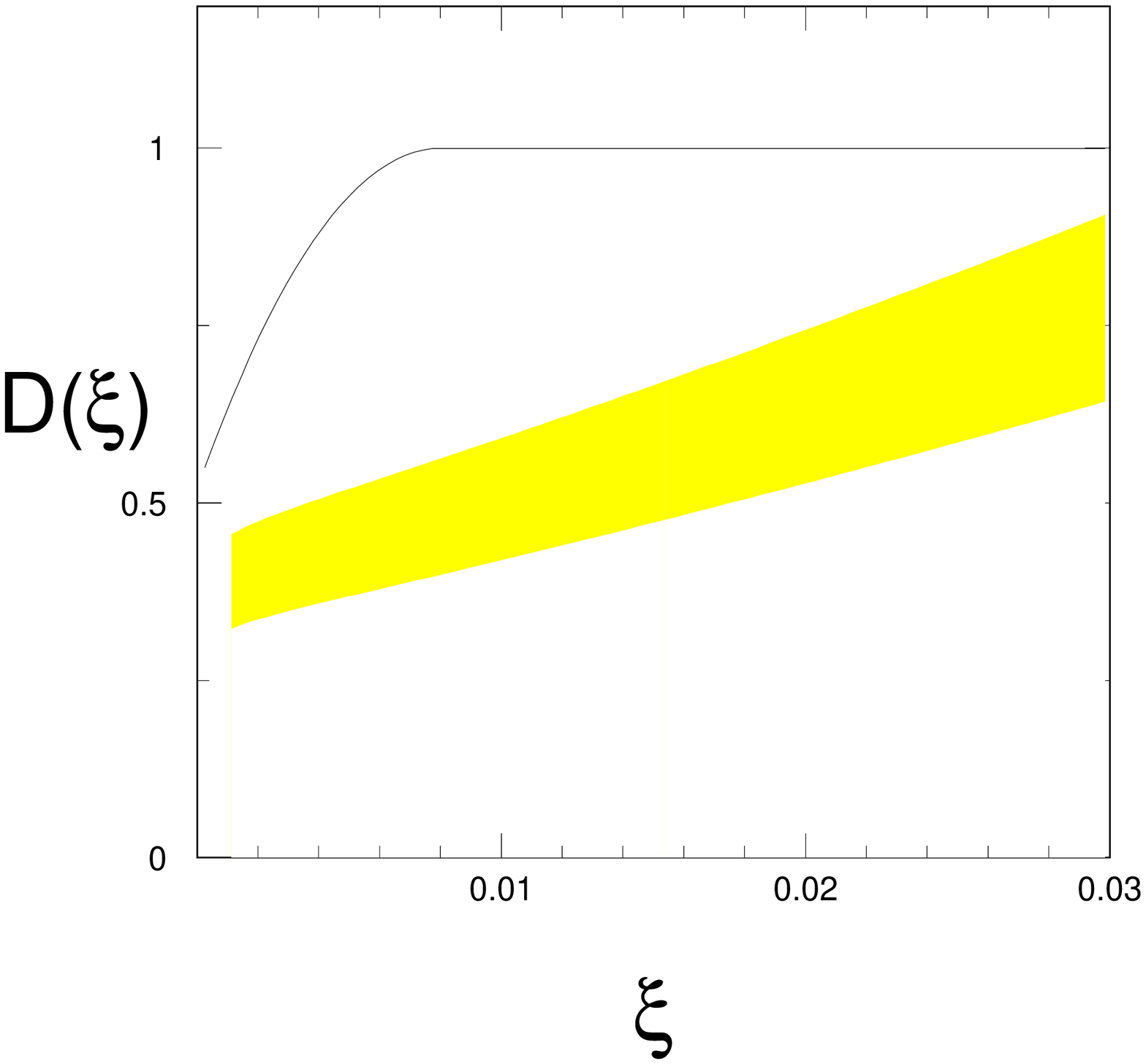,width=13cm}}
\end{center}
\caption[]{
The damping function referred to in the text. 
In the $\xi$--range, 0.0002--0.008, the function shown is a quadratic,
$D(\xi) = 1 - 7500 (0.008 - \xi)^2$. 
For $\xi < 0.0002$, the function is made to drop 
quickly to zero following a cubic function, as described in the text.
The band is the ratio of the CDF \dsig\
at $|t| = 0.05$~GeV$^2$ and $\sqrt{s} = 1800$~GeV
to the triple--Regge prediction described in the text.
}
\label{fig:damp}
\end{figure}

For $D(\xi)$, 
we have tried a ``toy'' damping function which decreases from unity for 
$\xi < 0.008$, following a quadratic function as shown in Fig.~\ref{fig:damp}.
The parameters of the quadratic function are choosen to reproduce the
leveling--off of \sigdiftot\ at ISR energies in Fig.~\ref{fig:sigtot}.
To account for the Tevatron and \SPS\ points, 
an additional (steep) fall--off is needed for $\xi < 0.0002$.
We arbitrarily use a cubic 
form\footnote{$D(\xi) = 7000 \xi - (2.86 \x 10^7) 
\xi^2 + (3.62 \x 10^{10}) \xi^3$.  
$D(\xi)$ has a total of 3 free parameters, since
the quadratic and cubic have identical slopes and magnitudes 
at $\xi = 0.0002$. These were chosen
to be the parameters of the quadratic function and the slope of the
cubic function at $\xi = 0$.}.
The dashed curves on Fig.~\ref{fig:sigtot} shows how such a 
function accounts reasonably 
well\footnote{The bump between the ISR and \SPS\ energies 
is due to the interplay between the 2--component \sigpomtot\ and
the damping function at $\xi_{min}$.}  
for \sigdiftot\ at high energies.

The solid curves in Figs.~\ref{fig:dsdtall} and \ref{fig:dsdtua8} are calculated
from Eq.~\ref{eq:tripleP2} multiplied by the above damping factor.
As expected from Fig.~\ref{fig:sigtot}, the effect of damping is very small
at ISR energies, but increases with $s$. At the energy of the \SPS , however,
the damping is about a factor of 3
in the low \T\ region, which dominates \sigdiftot .
There is good agreement with the damping predictions
at low--$|t|$ in both Figs.~\ref{fig:dsdtall} and \ref{fig:dsdtua8}. 

While, as explained above, the parameters of the quadratic term are chosen
to agree with the departure of the ISR cross sections from the triple-Regge
prediction in Fig.~\ref{fig:sigtot}, there seems to be no reason a-priori
why this formulation should be valid at higher energies. 
To clarify this point, we assume that the formula CDF fitted\cite{cdf}
to their data is a sufficient description of \dsig\ and compare its
$\xi$-dependence at $|t|$ = 0.05~GeV$^2$ with that of Eq.~\ref{eq:tripleP2}.
The band in Fig.~\ref{fig:damp} shows the ratio of the CDF \dsig\ at 
1800~GeV to the triple-Regge prescription, which 
can be interpreted as an empirical damping factor.
This decreases from near unity at $\xi = 0.03$ to about 0.5 at
$\xi = 0.01$, but is 
insufficient to account for the factor of $\sim$5 required by the 1800 GeV
cross section in Fig.~\ref{fig:sigtot}; 
therefore an additional (rapid) decrease in the 
damping factor must occur at smaller $\xi$, analogous to our toy damping 
factor discussed above.

The CDF function thus indicates 
that, at larger $s$, the onset of damping occurs at increasingly larger
$\xi$--values.
However, this does not invalidate the solid (damped) curve 
calculated with the above $D(\xi)$ and shown in
Fig.~\ref{fig:dsdtua8}, 
because the overall damping calculation is not
sensitive to the details of $D(\xi)$ in the larger $\xi$-region.

\section{Conclusions}
\indent

We first summarize the key points of this Letter:
\begin{itemize}
\item We have a triple-Regge parametrization of inclusive single diffraction
which agrees with the data in two domains of the $\xi$--\T\ plane:
(a) $\xi > 0.03$ at all \T , (b) $|t| > 1$~GeV$^2$ at all $\xi$
(Figs.~\ref{fig:dsdt035}, \ref{fig:sigtothit} and \ref{fig:cdft}).
Since the triple-Regge parametrization fails when applied to the full 
$\xi$--\T\ range of the total single diffractive cross section 
($\xi_{min} < \xi < 0.05$ and all \T ), we can conclude that
damping occurs only at low--$\xi$ and low--$|t|$.
\item We have given a parametrization of the damping factor, D($\xi$),
valid for $\T < 0.5$~GeV$^2$, which describes all the low--$|t|$ 
$d\sigdiftot /dt$ data at the ISR and
roughly accounts for the observed $s$--dependence of \sigdiftot\
up to Tevatron energies
(Figs.~\ref{fig:sigtot}, \ref{fig:dsdtall}, \ref{fig:dsdtua8}
and \ref{fig:damp}). 
\item  An effective damping factor calculated for the CDF fitted function 
for \dsig\ at $\sqrt{s} = 1800$~GeV and $|t| = 0.05$~GeV$^2$, 
suggests that, at fixed-$\xi$, damping increases as $s$ increases
(Fig.~\ref{fig:damp}).
\end{itemize}
These results raise a number of issues:
We can conjecture that, in the regions where the triple-Regge formalism
describes the data and there is no evidence of damping, 
factorization is valid and the \pom -flux-factor
may be universal.
A systematic program of testing universality
of the flux factor in these regions should be carried out
in $pp$, $\pap$ and $ep$ interactions.

Our damping results may shed some light on the measurements of the $fK$
quantity from the cross sections for diffractive jet production in
React.~\ref{eq:dif} and its analogue $ep$ reaction (K is the normalization
constant
in Eq.~\ref{eq:tripleP2} and $f$ measures violation of the momentum sum rule,
when $f \neq 1$). Assuming that the \pom\ has dominant gluonic 
structure\cite{h1gluonic}, there are three measurements of $fK$:

\vspace{4.0mm}
\begin{center}
\begin{tabular}{ll} 
UA8~\cite{jetsig}   &$fK = 0.30 \pm 0.10$ \\
CDF~\cite{cdfdijet} &$fK = 0.11 \pm 0.02$ \\
ZEUS~\cite{zeus}    &$fK = 0.37 \pm 0.15$ 
\end{tabular}
\end{center}
\vspace{4.0mm}

\noindent
The fact that the UA8 data is at large-$|t|$ where there is no damping, whereas
the CDF data is at small-$|t|$ where the damping factor in the region
of the jets is of order 0.50,
could account for the difference between the UA8 and CDF $fK$ values. 

In addition, despite the
large errors, it is interesting that the UA8 and ZEUS values for $fK$ are
consistent. This might be expected, if there is no damping in $ep$ collisions
at high-$Q^2$. Of course, at low-$Q^2$, where the photon exhibits hadronic
properties, multi-\pom\ exchange, and hence damping, may result in smaller
values of $fK$.

In order to further study the possible $s$-dependence of the effective damping
factor, it would be very useful to make detailed measurements of single
diffraction in pp interactions at RHIC energies. This would fill in the
large gap getween ISR and \SPS --collider 
energies seen in Fig.~\ref{fig:sigtot}.

We note that 
UA8\cite{ua8dif} offers as possible explanations of their observed
flattening of the \pom\ trajectory for $|t| > 1$~GeV$^2$, either that it is 
an effect of multiple-\pom\ exchange, or that it is 
evidence for the onset of the perturbative
2-gluon pomeron\cite{fs,cfs}. 
In view of our observation that damping is not required
in this \T -region, it seems that the perturbative \pom , 
explanation is more likely.
It may therefore be interesting to study the $\gamma$--\pom\ cross section
from $t_{min}$ up through the $|t| > 1$~GeV$^2$ region.

\section*{Acknowledgements}
\indent

We thank A. Kaidalov, P. Landshoff and E. Gotsman for helpful 
comments.

\pagebreak

\clearpage
 
\clearpage
 
\clearpage

\clearpage
 
\clearpage

\clearpage

\clearpage
 
\begin{thebibliography}{99}

\bibitem{reviews}
For reviews, see e.g.:\\
\noindent
A.B.~Kaidalov, Phys. Reports 50 (1979) 157;\\
G. Alberi and G. Goggi, Phys. Reports 74 (1981) 1;\\
K. Goulianos, Phys. Reports 101 (1983) 169.

\bibitem{is}
G. Ingelman and P.E. Schlein, Phys. Lett. B 152 (1985) 256.

\bibitem{bonino}
R. Bonino et al. (UA8 Collaboration), Phys. Lett. B 211 (1988) 239.

\bibitem{brandt}
A. Brandt et al. (UA8 Collaboration), Phys. Lett. B 297 (1992) 417.

\bibitem{jetsig}
A. Brandt et al. (UA8 Collaboration),  Phys. Lett. B 421 (1998) 395.

\bibitem{dino}
K. Goulianos, Phys. Lett. B358 (1995) 379; B363 (1995) 268.

\bibitem{mueller}
A.H.~Mueller, Phys. Rev. D2 (1970) 2963; D4 (1971) 150;\\
A.B.~Kaidalov et al., Pisma JETP 17 (1973) 626;\\
A.~Capella, Phys. Rev. D8 (1973) 2047;\\
R.D.~Field and G.C.~Fox, Nucl. Phys. B80 (1974) 367;\\
D.P.~Roy and R.G.~Roberts, Nucl. Phys. B77 (1974) 240.

\bibitem{bc24}
V. Blobel et al., Nucl. Phys. B92 (1975) 221.

\bibitem{bc69}
H.~Bialkowska et al., Nucl. Phys. B110 (1976) 300.

\bibitem{bc100}
J.W.~Chapman et al., Phys. Rev. Lett. 32 (1974) 257;
the cross section at 405 GeV is multiplied by a factor 0.82 to estimate
the value for $x_p > 0.95$.

\bibitem{bc200}
S.J.~Barish et al., Phys. Rev D9 (1974) 2689; Phys. Rev. Lett. 31 (1973) 1080.

\bibitem{bc300}
F.T.~Dao et al., Phys. Lett. B45 (1973) 399.

\bibitem{albrowtot}
M.G. Albrow et al., Nucl. Phys. B108 (1976) 1.

\bibitem{armitagetot}
J.C.M. Armitage et al., Nucl. Phys. B194 (1982) 365.

\bibitem{ua4dif1}
M. Bozzo et al. (UA4 Collaboration), Phys. Lett. B136 (1984) 217.

\bibitem{ua4dif2}
D. Bernard et al. (UA4 Collaboration), Phys. Lett. B186 (1987) 227.

\bibitem{cdf}
F.~Abe et al. (CDF Collaboration), Phys. Rev. D50 (1994) 5535.

\bibitem{theory1}
A.~Capella, J.~Kaplan and J.~Tran Thanh Van, Nucl. Phys. B 105 (1976) 333.

\bibitem{theory2}
A.B.~Kaidalov, L.A.~Ponomarev and K.A.~Ter--Martirosyan, 
Sov. J. Nucl. Phys. 44 (1986) 468.

\bibitem{theory3}
P.~Aurenche et al., Phys. Rev. D 45 (1992) 92.

\bibitem{theory4}
E.~Gotsman, E.M.~Levin and U.~Maor, Zeit. Phys. C 57 (1993) 667; 
Phys. Rev. D 49 (1994) R4321; Phys. Lett B 353 (1995) 526.

\bibitem{theory5}
A.~Capella, A.~Kaidalov, C.~Merino and J.~Tran Thanh Van, 
Phys. Lett. B 337 (1994) 358.

\bibitem{kaidalov}
A.~Kaidalov, private communication (1998).

\bibitem{dl_elastic}
A. Donnachie \& P.V. Landshoff, Nucl. Phys. B231 (1984) 189; 
Nucl. Phys. B267 (1986) 690.

\bibitem{ua8dif}
A. Brandt et al. (UA8 Collaboration), Nucl. Phys. B 514 (1998) 3.

\bibitem{dl_tot}
A. Donnachie \& P.V. Landshoff, Phys. Lett. B296 (1992) 227.

\bibitem{dl_dif}
A. Donnachie \& P.V. Landshoff, Nucl. Phys. B244 (1984) 322.

\bibitem{cudell}
J.R.~Cudell, K.~Kyungsik and K.K.~Sung, Phys. Lett. B395 (1997) 311;\\
J.R. Cudell, K. Kang and S.K. Kim, ``Simple Model for Total Cross Sections",
preprint, Brown--HET--1060, January 1997.

\bibitem{dino2}
R.J.M.~Covolan, J.~Montanha \& K.~Goulianos, Phys. Lett. B389 (1996) 176.

\bibitem{albrow}
M.G. Albrow et al., Nucl. Phys. B54 (1973) 6; \\
M.G. Albrow et al., Nucl. Phys. B72 (1974) 376.

 \bibitem{h1gluonic}
C.~Adloff et al. (H1 Collaboration), Z. Phys. C76 (1997) 613.

\bibitem{cdfdijet}
F.~Abe et al. (CDF Collaboration), Phys. Rev. Lett. 79 (1997) 2636.

\bibitem{zeus}
M. Derrick et al. (ZEUS Collaboration), Phys. Lett. 356 (1995) 129.

\bibitem{fs}
L. Frankfurt and M. Strikman, Phys. Rev. Lett. 63, (1989) 1914;
64 (1990) 815.

\bibitem{cfs}
J.C.~Collins, L.~Frankfurt and M.~Strikman, Phys. Lett. B307 (1993) 161.


\end{thebibliography}
\end{document}